\documentclass[aps,prd,preprint,superscriptaddress,tightenlines,nofootinbib,showpacs]{revtex4}

\usepackage{graphicx}
\usepackage{dcolumn}
\usepackage{bm}
\usepackage{amsmath}
\usepackage{amssymb}

\def\ee{\ifmmode e^+e^- \else $e^+e^-$  \fi}
\def\mm{\ifmmode \mu^+\mu^- \else $\mu^+\mu^-$  \fi}
\def\LL{\ifmmode l^+l^- \else $l^+l^-$  \fi}

\begin{document}

\preprint{CLNS 05/1939}       
\preprint{CLEO 05-27}         

\title{Measurement of 
$\sigma(
e^+e^-\to\psi(3770)\to\mbox{hadrons})$ 
at E$_{\mbox{\scriptsize{cm}}}=3773$~MeV}


\author{D.~Besson}
\affiliation{University of Kansas, Lawrence, Kansas 66045}
\author{T.~K.~Pedlar}
\affiliation{Luther College, Decorah, Iowa 52101}
\author{D.~Cronin-Hennessy}
\author{K.~Y.~Gao}
\author{D.~T.~Gong}
\author{J.~Hietala}
\author{Y.~Kubota}
\author{T.~Klein}
\author{B.~W.~Lang}
\author{R.~Poling}
\author{A.~W.~Scott}
\author{A.~Smith}
\affiliation{University of Minnesota, Minneapolis, Minnesota 55455}
% ...
\collaboration{CLEO Collaboration} 
\noaffiliation


\date{December 15, 2005}

\begin{abstract} 
  We measure the cross section for 
$e^+e^-\to\psi(3770)\to\mbox{hadrons}$ at
E$_{\mbox{\scriptsize{cm}}}=3773$~MeV
to be $(6.38\pm0.08^{+0.41}_{-0.30})$~nb using the CLEO detector 
at the CESR $e^+e^-$ collider.
The difference between this and the 
$e^+e^-\to\psi(3770)\to D\bar{D}$ cross section at the
same energy is found to be $(-0.01\pm0.08^{+0.41}_{-0.30})$~nb.
With the observed total cross section,
we extract $\Gamma_{ee}(\psi(3770))=(0.204\pm0.003^{+0.041}_{-0.027})$~keV.
 Uncertainties shown are statistical
and systematic, respectively.
\end{abstract}

\maketitle

 Two decades ago, the Mark III collaboration~\cite{mark3} measured the 
 cross section for the reaction $e^+e^-\to\psi(3770)\to D\bar{D}$, 
 using a double-tag technique.
 They found a cross section of about 5 nb.  At roughly the same time, 
 the Mark II~\cite{mark2}
 and Lead-Glass Wall~\cite{leadgas} collaborations 
 measured the cross section for
 $e^+e^-\to\psi(3770)\to\mbox{hadrons}$ (including $D\bar{D}$), 
 finding a cross section of about 10 nb.  The indication that the decay of 
 $\psi(3770)$  to non-$D\bar{D}$ final states
 was comparable to that to $D\bar{D}$ final states came as a surprise.  The
 expectation was that the decay width of $\psi(3770)$ to non-$D\bar{D}$ 
 final states would be comparable to that of $\psi(2S)$,
 which is below threshold for open charm decays.
 Thus $\mathcal{B}(\psi(3770)\to$ non-$D\bar{D})/\mathcal{B}(\psi(3770)\to
 D\bar{D})$ should be small
 due to the large total width
 of $\psi(3770)$.
 The CLEO collaboration has recently measured the cross 
 section for
 $e^+e^-\to \psi(3770)\to D\bar{D}$, 
 finding $\sigma_{\psi(3770)\to D\bar{D}}=(6.39\pm0.10^{+0.17}_{-0.08})$~nb
 at E$_{\mbox{\scriptsize{cm}}}=3773$~MeV~\cite{xsecddcleo}.
 In this letter
 we present a measurement of the cross section for
 $e^+e^-\to\psi(3770)\to\mbox{hadrons}$, $\sigma_{\psi(3770)}$, at 
 E$_{\mbox{\scriptsize{cm}}}=3773$~MeV, where $\psi(3770)$ refers to 
 the yield at E$_{\mbox{\scriptsize{cm}}}=3773$~MeV from $c\bar{c}$
 annihilation into
 hadrons, not including continuum production of $q\bar{q}$ ($q=u,d,s$) and
 not including radiative returns to $\psi(2S)$ and to $J/\psi$.

 We define $\sigma_{\psi(3770)}$ as
 \begin{gather}\label{eq:1}
   \sigma_{\psi(3770)}=\frac{N_{\psi(3770)}}
	 {\epsilon_{\psi(3770)}\cdot\mathcal{L}_{\psi(3770)}},
 \end{gather}
 \noindent where
 $\mathcal{L}_{\psi(3770)}$ is the integrated luminosity for the data taken
 at E$_{\mbox{\scriptsize{cm}}}=3773$~MeV,
 $N_{\psi(3770)}$ is the number of hadronic events 
 inferred to be directly from
 $\psi(3770)$ decays, and
 $\epsilon_{\psi(3770)}$ is the hadronic event selection efficiency of
 $\psi(3770)$ decays.

 Our main observable is the background subtracted 
 number of hadrons produced in 
 $\psi(3770)$ decays, $N_{\psi(3770)}$.
 At E$_{\mbox{\scriptsize{cm}}}\sim 3773$~MeV, 
 the main backgrounds come from continuum
 production $e^+e^-\to q\bar{q}$ and  radiative returns to 
 $\psi(2S)$ and $J/\psi$. Thus $N_{\psi(3770)}$ is given  by
 \begin{gather}\label{eq:4}
   N_{\psi(3770)}=N_{\mbox{\scriptsize{on-}}\psi(3770)}-N_{q\bar{q}}-
   N_{\psi(2S)} 
   - N_{J/\psi} - \Sigma_{l=\tau,\mu,e}N_{\ell^+\ell^-},
 \end{gather}
 \noindent where $N_{\mbox{\scriptsize{on-}}\psi(3770)}$ 
 is the observed number of hadronic
 events in the $\psi(3770)$ data taken at 
 E$_{\mbox{\scriptsize{cm}}}=3773$~MeV,
 $N_{q\bar{q}}$ is the number of observed hadronic events from 
 $e^+e^-\to\gamma^*\to q\bar{q}$, 
 $N_{\psi(2S)}$ and $N_{J/\psi}$ are the
 number of hadronic events from $\psi(2S)$ and $J/\psi$ decays respectively,
 and
 $N_{\ell^+\ell^-}$ is the number of events from 
 $e^+e^-\to \ell^+\ell^-$ that pass our
 hadronic event selection criteria. We subtract these backgrounds by employing
 scaled numbers of hadrons observed in two other data samples, taken at
 the $\psi(2S)$ peak (E$_{\mbox{\scriptsize{cm}}}=3686$~MeV) and at
 the continuum below this resonance (E$_{\mbox{\scriptsize{cm}}}=3671$~MeV).

  The three $e^+e^-$ collision data samples 
  taken at
  E$_{\mbox{\scriptsize{cm}}}=3671$, $3686$, and $3773$~MeV
  were acquired with
  the $\mbox{CLEO-c}$ detector~\cite{cleodet}
  operating at the Cornell Electron Storage Ring \cite{cesr},
  corresponding to integrated luminosities of 
  $\mathcal{L}=(20.7\pm0.2)\mbox{pb}^{-1}$, $(2.9\pm0.1)\mbox{pb}^{-1}$, and 
  $(281.3\pm2.8)\mbox{pb}^{-1}$, respectively.
  Components of the $\mbox{CLEO-c}$ detector used for this analysis are 
  the charged particle tracking system
  (the drift chamber)
  operating in a 1.0~T magnetic field along the beam axis and achieving
  a momentum resolution of $\sim 0.6\%$ at momenta of $1$~GeV/c, and the
  CsI crystal calorimeter attaining photon energy resolution of $2.2\%$
  for $E_\gamma = 1$~GeV and $5\%$ at $100$~MeV. Together, they cover $93\%$
  of solid angle for charged and neutral particles. 
  The RICH detector and muon system are not used for this analysis.

  To select hadronic events, we require that the observed number of charged
  tracks (N$_{\mbox{\scriptsize{ch}}}$) be at least three. 
  The tracks are required to have well-measured momenta and to satisfy
  criteria based on track fit quality. They must also be consistent with
  originating from the interaction point in three dimensions (we vary these
  track quality requirements for study of systematic errors).
  The visible energy of charged and neutral showers
  (E$_{\mbox{\scriptsize{vis}}}$) must be at least $30\%$ of the 
  center-of-mass energy 
  (E$_{\mbox{\scriptsize{cm}}}$).
  For $3\leq$ N$_{\mbox{\scriptsize{ch}}}\leq 4$, 
  the total energy visible in the calorimeter alone (E$_{\mbox{\scriptsize{cal}}}$)
  must be at least $15\%$ of E$_{\mbox{\scriptsize{cm}}}$ and, to suppress 
  $e^+e^-\to e^+e^-$, 
  the most energetic shower in the calorimeter
  must be less than $75\%$ of the beam energy or E$_{\mbox{\scriptsize{cal}}}
  <0.85~$E$_{\mbox{\scriptsize{cm}}}$. 

  Some remaining backgrounds can be  virtually eliminated with further 
  restrictions.
  Two-photon fusion events 
  ($e^+e^-\to e^+e^-\gamma^*\gamma^*\to e^+e^-+\mbox{hadrons}$) 
  can be reduced to a negligible
  background by requiring there to be no large momentum imbalance
  along the beam direction ($z$-axis). To accomplish this, we require
  that the ratio of the $z$-component of the vector sum of all charged
  particles and photon candidates to the visible energy, 
  $|\vec{P}_z^
  {\mbox{\scriptsize{net}}}|/$E$_{\mbox{\scriptsize{vis}}}$,
  be less than 0.3. Monte Carlo (MC) studies show that this
  selection causes a loss of only $\sim 4\%$ of signal events. 
  Backgrounds from
  cosmic rays and collisions of beam particles with gas molecules
  or the walls of the vacuum pipe are suppressed by restrictions
  on the event vertex, defined as the average of the intersection
  points of all charged track pairs in an event. True $e^+e^-$ collision 
  events will peak sharply near the collision point, with rms
  widths of $\sim 1$~mm in the $xy$ plane and $\sim 1$~cm along the $z$-axis.
  We require the vertex be closer than 5~mm in the $xy$ plane 
  and 5~cm along the $z$-axis, which leave just a few tenths
  percent backgrounds from these sources.

  The efficiency of the event selection criteria is $\sim 80\%$ 
  for $D\bar{D}$ events,
  which, as will be shown, constitute the bulk of $\psi(3770)$ decays.
  The
  absolute efficiencies for continuum $q\bar{q}\to\mbox{hadrons}$
  and for radiative
  returns to $\psi(2S)$ are smaller, but we need not expose the analysis
  to uncertainties in modeling these processes, because we are
  able to use data samples acquired with the same detector at
  lower energies for these major subtractions. A large source
  of potential systematic uncertainty is avoided with this strategy.

  $N_{J/\psi}$ and $N_{\ell^+\ell^-}$ in Eq.~\ref{eq:4} are obtained by
  $N_{J/\psi}=\sigma_{e^+e^-\to\gamma J/\psi}\cdot\mathcal{L}_{\psi(3770)}\cdot
  \epsilon_{e^+e^-\to\gamma J/\psi}$ and $N_{\ell^+\ell^-}
  =\sigma_{e^+e^-\to \ell^+\ell^-}
  \cdot\mathcal{L}_{\psi(3770)}\cdot\epsilon_{e^+e^-\to 
    \ell^+\ell^-}$ respectively,
  where the production cross sections,
  $\sigma_{e^+e^-\to\gamma J/\psi}$ and $\sigma_{e^+e^-\to \ell^+\ell^-}$, are
  theoretically estimated.
  In particular, $\sigma_{e^+e^-\to\gamma J/\psi}$
  is calculated based on
  the radiative tail kernel \cite{radker} convoluted with the
  resonance Breit-Wigner shape with continuum interference (will be 
  discussed later in this report).
  $\epsilon_{e^+e^-\to\gamma J/\psi}$ and $\epsilon_{e^+e^-\to \ell^+\ell^-}$
  are the hadronic event selection efficiencies of
  events from
  radiative return to $J/\psi$ and from $e^+e^-\to \ell^+\ell^-$
  respectively,
  determined by the EvtGen event generator 
  \cite{evtgen} and a GEANT-based detector simulation \cite{geant321}.

  $N_{\psi(2S)}$ in Eq.~\ref{eq:4} is given by
  $ N_{\psi(2S)}=R_{\pi\pi\ell\ell}\cdot N_{\psi(2S)}(3686)$, where
  $R_{\pi\pi\ell\ell}=0.59\pm0.01$~(stat.)
  is the ratio of observed numbers of $\pi^+\pi^-\ell^+\ell^-$ events at 
  E$_{\mbox{\scriptsize{cm}}}=3773$~MeV \cite{pipi3s} to 
  that at $3686$~MeV \cite{pipi2s}, $\ell^\pm\equiv e^\pm\mbox{ or }\mu^\pm$,
  and $N_{\psi(2S)}(3686)$ is the
  observed number of hadronic events ($(1,019\pm1)\times 10^3$)
  in data taken at E$_{\mbox{\scriptsize{cm}}}=3686$~MeV
  after subtracting continuum backgrounds ($(34.0\pm0.2)\times 10^3$).
  To estimate the continuum contribution we scale the 
  E$_{\mbox{\scriptsize{cm}}}=3671$~MeV
  hadronic yield by the ratio of integrated luminosities corrected for
  $1/s$ dependence of the continuum processes.
  Since this subtraction is small, $\psi(2S)$ contribution to the
  E$_{\mbox{\scriptsize{cm}}}=3671$~MeV 
  data and different $s$-dependence of the
  backgrounds from $e^+e^-\to\tau^+\tau^-$ and $e^+e^-\to\gamma J/\psi$
  events can be safely neglected.

  To obtain the largest background at E$_{\mbox{\scriptsize{cm}}}=3773$~MeV,
  $N_{q\bar{q}}$ in Eq.~\ref{eq:4}, we
  employ the data taken at E$_{\mbox{\scriptsize{cm}}}=3671$~MeV,
  which has small contaminations
  of $\psi(2S)$ decays as well as of $J/\psi$ decays. Hence we obtain
  $N_{q\bar{q}}$ by
  \begin{gather}
  N_{q\bar{q}}=S\cdot N_{q\bar{q}}(3671)
  =S\cdot\{N_{\mbox{\scriptsize{had}}}(3671)-N_{\psi(2S)}(3671) 
    -N_{J/\psi}(3671)-\Sigma_{l=\tau,\mu,e}N_{l^+l^-}(3671)\},
  \end{gather}
  \noindent where all $N(3671)$'s are the number of hadronic events
  at E$_{\mbox{\scriptsize{cm}}}=3671$~MeV. $N_{\psi(2S)}(3671)$,
  $N_{J/\psi}(3671)$, and $N_{l^+l^-}(3671)$ are all obtained in the same
  way as described previously but at E$_{\mbox{\scriptsize{cm}}}=3671$~MeV. 
  The scaling factor,
  $S=12.88\pm0.01$~(stat.),
  accounts for the luminosity difference between the two data sets and for
  the $1/s$ dependence of the cross section.
  
  In addition to the above corrections, 
  we must take into account the effect of interference between 
  the final states of resonance decays (i.e. 
  $\mbox{resonance}\to\gamma^*\to q\bar{q}\to\mbox{hadrons}$) and 
  non-resonant annihilation of 
  $e^+e^-$ (i.e. $e^+e^-\to q\bar{q}\to\mbox{hadrons}$) 
  as it distorts the shape and area of the intrinsic 
  Breit-Wigner line shape.
  To estimate the size of this effect, we assume the
  following:
  \begin{itemize}
    \item The amplitude for $\gamma^*\to q\bar{q}$ interferes
      in the same way as the one for $\gamma^*\to\mu^+\mu^-$.
    \item 
      We can treat
      $ggg\to$ hadrons and 
      $q\bar{q}\to$ hadrons as distinct final states. 
      Thus they are incoherent and do not interfere with each other.
  \end{itemize}
  The interference effects due to decay products of $J/\psi$ should be
  negligible compared to that of $\psi(2S)$.
  They are taken into account by the subtraction of the scaled
  continuum data.
  With the first assumption, 
  the change in cross section of
  $e^+e^-\to\gamma^*\to q\bar{q}$ due to 
  the interference, $\sigma_{q\bar{q}}^{\mbox{\scriptsize{inter}}}$,
  is given by
  \begin{gather}\label{eq:qqhad}
    \sigma_{q\bar{q}}^{\mbox{\scriptsize{inter}}}
    =\frac{R}{1-2{\cal B}_{\psi(2S)\to\mu^+\mu^-}}
    \cdot \sigma_{\mu\mu}^{\mbox{\scriptsize{inter}}},
  \end{gather}
  \noindent where $\sigma_{\mu\mu}^{\mbox{\scriptsize{inter}}}$
  is the change in cross section
  due to the interference between $\psi(2S)\to\gamma^*\to\mu^+\mu^-$ and
  $e^+e^-\to\gamma^*\to\mu^+\mu^-$~\cite{inter,inter-2}.
  We take a $\pm 25\%$ uncertainty in 
  $\sigma_{q\bar{q}}^{\mbox{\scriptsize{inter}}}$ as 
  an estimate
  of the systematic error from this term.
  $R$ is 
  $\sigma_{q\bar{q}\to\mbox{\scriptsize{hadrons}}}/\sigma_{\mu\mu}^0$, 
  which can be also written as 
  $N_{q\bar{q}}^{\mbox{\scriptsize{corr}}}/(\mathcal{L}\cdot\epsilon_{q\bar{q}}
  \cdot\sigma_{\mu\mu}^0)$, where 
  $\sigma_{\mu\mu}^0$ is the lowest-order
  $\mu^+\mu^-$ cross section,
  $N_{q\bar{q}}^{\mbox{\scriptsize{corr}}}$ is the number
  of hadronic event from $e^+e^-\to\gamma^*\to q\bar{q}$, corrected for
  this interference effect, 
  $\mathcal{L}$ is the 
  integrated luminosity, and $\epsilon_{q\bar{q}}$
  is the event selection efficiency of $e^+e^-\to\gamma^*\to q\bar{q}\to
  \mbox{hadrons}$.

  We obtain the final $N_{q\bar{q}}$  by
  correcting for the \textit{destructive} and \textit{constructive}
  interference effects 
  at E$_{\mbox{\scriptsize{cm}}}=3671$~MeV and $3773$~MeV respectively.
  We also account for the interference effects between
  $e^+e^-\to\gamma^*\to\ell^+\ell^-$ in continuum and 
  $\psi(2S)\to\gamma^*\to\ell^+\ell^-$, 
  but these effects are negligible compared to
  the one described above.
  The corrections for resonance-continuum $q\bar{q}$ interference in our 
  data samples
  amount to an $11\%$ downward shift in the cross section at
  E$_{\mbox{\scriptsize{cm}}}=3773$~MeV.

  Tables~\ref{tab:num} and \ref{tab:numres} present the observed 
  numbers of hadronic events 
  (only statistical errors are shown)
  from various specific sources in the two data samples
  taken
  at E$_{\mbox{\scriptsize{cm}}}=3671$~MeV and at $3773$~MeV, respectively.
  \begin{table}
  \caption{Numbers of events in $10^3$ at 
    E$_{\mbox{\scriptsize{cm}}}=3671$~MeV
    for various event types.
    The interference term represents the interference of
    $\psi(2S)\to\gamma^*\to q\bar{q}\to\mbox{hadrons}$ with
    the continuum annihilation, $\gamma^*\to q\bar{q}\to\mbox{hadrons}$.
  \label{tab:num}}
  \def\1#1#2#3{\multicolumn{#1}{#2}{#3}}
  \begin{center}
  \begin{tabular}{l|r}
  \hline
  \hline
  $N_{\mbox{\scriptsize{had}}}(3671)$       & $244.4\pm0.5$ \\
  $N_{\psi(2S)}(3671)$  & $7.2\pm0.5$ \\
  $N_{J/\psi}(3671)$    & $13.0\pm0.1$ \\
  $N_{\tau^+\tau^-}(3671)$        & $9.1\pm0.1$ \\
  $N_{e^+e^-}(3671)$              & $2.7\pm0.5$ \\
  $N_{\mu^+\mu^-}(3671)$          & $0.13\pm0.02$ \\
  Interference (\textit{destructive})& $8.9\pm0.0$\\
  \hline
  $N_{q\bar{q}}^{\mbox{\scriptsize{corr}}}$  & $221\pm1$ \\
  \hline
  \hline
  \end{tabular}
  \end{center}
  \end{table}
  
  \begin{table}
    \caption{Numbers of events in $10^3$ at 
    E$_{\mbox{\scriptsize{cm}}}=3773$~MeV
    for various event types.
    The interference term represents the interference of
    $\psi(2S)\to\gamma^*\to q\bar{q}\to\mbox{hadrons}$ with
    the continuum annihilation, $\gamma^*\to q\bar{q}\to\mbox{hadrons}$.
    The scaled $N_{q\bar{q}}$ is raised by $\sim 2\%$ to correct for
    the difference in efficiencies ($\epsilon_{q\bar{q}}$)
    at the two energy points,
    $3671$~MeV and $3773$~MeV.
    \label{tab:numres}}
    \def\1#1#2#3{\multicolumn{#1}{#2}{#3}}
  \begin{center}
  \begin{tabular}{l|r}
  \hline
  \hline
  $N_{\mbox{\scriptsize{on-}}\psi(3770)}$   & $5,319\pm2$ \\
  $N_{q\bar{q}}$        & $2,915\pm12$ \\
  Interference (\textit{constructive}) & $26.8\pm0.1$ \\
  $N_{\psi(2S)}$  & $583\pm6$ \\
  $N_{J/\psi}$    & $140\pm1$ \\
  $N_{\tau^+\tau^-}$        & $170.9\pm0.3$ \\
  $N_{e^+e^-}$              & $54\pm8$ \\
  $N_{\mu^+\mu^-}$          & $2.0\pm0.2$ \\
  \hline
 $N_{\psi(3770)}$ & $1,427\pm16$ \\
  \hline
  \hline
  \end{tabular}
  \end{center}
  \end{table}
  In Eq.~\ref{eq:1},
  the hadronic event selection efficiency, $\epsilon_{\psi(3770)}$, is expected
  to be close to $\epsilon_{D\bar{D}}$. 
  To account for the uncertainty of 
  $\epsilon_{\mbox{\scriptsize{non}}-D\bar{D}}$,
  the hadronic event selection efficiency of non-$D\bar{D}$ decays of 
  $\psi(3770)$,
  we include the non-$D\bar{D}$ component in the calculation of
  $\sigma_{\psi(3770)}$,
  using the formula
  \begin{gather}
    \sigma_{\psi(3770)}=\left(\frac{N_{\psi(3770)}}
    {\epsilon_{D\bar{D}}\cdot\mathcal{L}_{\psi(3770)}} - 
    \sigma_{\psi(3770)\to D\bar{D}}\right)\cdot
    \frac{\epsilon_{D\bar{D}}}{\epsilon_{\mbox{\scriptsize{non}}-D\bar{D}}}
    +\sigma_{\psi(3770)\to D\bar{D}}.
  \end{gather}
  \noindent We use
  CLEO's measurement~\cite{xsecddcleo} for $\sigma_{\psi(3770)\to D\bar{D}}$
  and assume $\epsilon_{\mbox{\scriptsize{non}}-D\bar{D}}$ 
  is the average of the efficiency of 
  $D\bar{D}$ decays ($\sim 80\%$) and that of $\psi(2S)$ ($\sim 68\%$). 
  We vary the efficiency between these
  two extremes to account for the uncertainty of
  $\epsilon_{\mbox{\scriptsize{non}}-D\bar{D}}$. However, as the difference
  between $\sigma_{\psi(3770)\to D\bar{D}}$ and $\sigma_{\psi(3770)}$
  turns out to be small as we will show shortly, the final 
  $\sigma_{\psi(3770)}$ becomes rather insensitive to 
  $\epsilon_{\mbox{\scriptsize{non}}-D\bar{D}}$
  and is more sensitive to the input value of 
  $\sigma_{\psi(3770)\to D\bar{D}}$.

\begin{figure}
    \includegraphics*[width=3.4in]{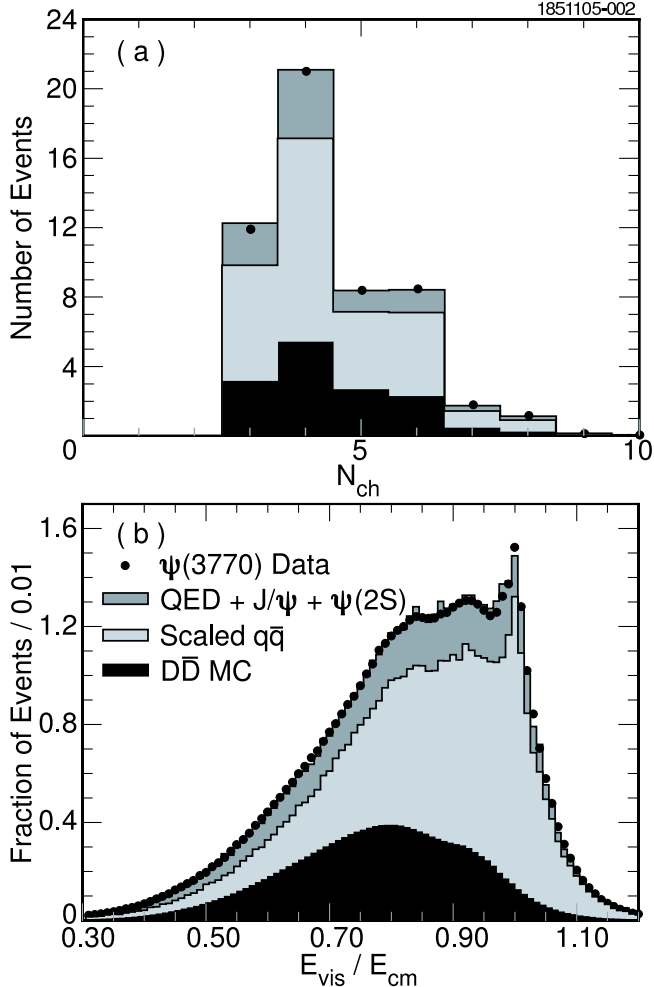}
  \caption{
  N$_{\mbox{\scriptsize{ch}}}$ (a:top) and
  E$_{\mbox{\scriptsize{vis}}}$/E$_{\mbox{\scriptsize{cm}}}$ 
  (b:bottom)
  of our $\psi(3770)$ sample that pass our hadronic event selection criteria 
  (black-solid histograms). 
  Backgrounds are also overlaid
  (generic $D\bar{D}$ Monte Carlo~\cite{evtgen,geant321}, 
  scaled continuum ($q\bar{q}$)
  data, summed QED events (
  $\Sigma_{l=e, \mu, \tau} (e^+e^-\to\ell^+\ell^-)$) plus
  radiative returns to $\psi(2S)$  and $J/\psi$).
   The yield of $D\bar{D}$
  Monte Carlo is scaled to the size of the data assuming
  $\sigma(e^+e^-\to D\bar{D}\to\mbox{hadrons})=6.4$~nb.}
  \label{fig:evtvar}
  \end{figure}

  Figure~\ref{fig:evtvar} shows the distributions of track 
  multiplicity (top) and visible
  energy 
  normalized to E$_{\mbox{\scriptsize{cm}}}$
  (bottom)
  of events in our $\psi(3770)$ data sample 
  that pass our hadronic event selection criteria 
  (black-solid histograms). 
  Also overlaid are various estimated and observed
  backgrounds.

  The total fractional systematic uncertainty in $\sigma_{\psi(3770)}$ 
  is $^{+6.5}_{-4.7}\%$,
  which is the quadrature sum of the fractional uncertainties due to various
  sources shown in Table \ref{tab:3770systsum}.
  One of the major sources of systematic error is 
  the accuracy of Monte Carlo modeling of those event characteristics that
  are used in event selection.
  We vary some of our event selection criteria,
  particularly the charged track multiplicity, 
  N$_{\mbox{\scriptsize{ch}}}$, and see
  the effect on our final $\sigma_{\psi(3770)}$.
  \begin{table}
  \caption{Summary of various relative systematic uncertainties 
    for hadronic cross section
  of $\psi(3770)$.
  \label{tab:3770systsum}}
  \begin{center}
  \def\1#1#2#3{\multicolumn{#1}{#2}{#3}}
  \begin{tabular}{l c c}
  \hline
  \hline
    source of error                          & $(\%)$ 
                                             & (nb) \\
  \hline
   Two-Photon suppression                    & $0.1$ & $0.01$ \\
   BeamGas/Wall/Cosmic subtraction           & $0.6$ & $0.04$ \\
   $N_{\psi(2S)}$                            & $0.9$ & $0.06$ \\
   $N_{J/\psi}$                              & $0.8$ & $0.05$ \\
   Track quality cuts                        & $0.6$ & $0.04$ \\
   Luminosity                                & $1.1$ & $0.07$ \\
   Continuum scaling                         & $2.2$ & $0.14$ \\
   Hadronic event selection criteria         & $^{+4.7}_{-1.4}$ 
                                             & $^{+0.30}_{-0.09}$ \\
   $\sigma_{\psi(3770)\to D\bar{D}}$         & $^{+0.1}_{-0.2}$ & $0.01$ \\
   $\epsilon_{D\bar{D}}$  & $1.7$ & $0.11$ \\
   Ratio of $q\bar{q}$ efficiencies          & $0.9$ & $0.06$ \\
   Interference ($q\bar{q}\to$ hadrons)      & $2.8$ & $0.18$ \\
   \hline
   Total                                     & $^{+6.5}_{-4.7}$
                                             & $^{+0.41}_{-0.30}$ \\
  \hline
  \hline
  \end{tabular}
  \end{center}
  \end{table}
  The uncertainty in the estimation of number of $\psi(2S)$ in on-resonance 
  data
  comes mainly from a small difference in signal efficiency of selecting
  $\pi^+\pi^-\ell^+\ell^-$ events between the two data sets 
  (E$_{\mbox{\scriptsize{cm}}}=3686$~MeV and $3773$~MeV).
  To estimate possible systematic variation due to hadronic
  event selection efficiency of generic decay of $D\bar{D}$, we also explore
  changes to the decay branching ratios of $D$ mesons used in the $D\bar{D}$
  Monte Carlo simulation when we vary track 
  multiplicities while observing changes in number of reconstructed $D\bar{D}$.
  Based on this study, we
  conservatively assign $1.7\%$ as an uncertainty in $\sigma_{\psi(3770)}$.
  
  The final cross section, including systematic uncertainty, is
  $\sigma_{\psi(3770)}=(6.38\pm0.08^{+0.41}_{-0.30})$~nb, where
  the first error is statistical and the second error is systematic.
  The difference between 
  $\sigma_{\psi(3770)\to D\bar{D}}$~\cite{xsecddcleo} 
  and $\sigma_{\psi(3770)}$ is $(-0.01\pm0.08^{+0.41}_{-0.30})$~nb,
  which is consistent with
  recently observed non-$D\bar{D}$ decays of $\psi(3770)$ 
  \cite{pipi3s, chic13s}.

  In addition to the measurement of $\sigma_{\psi(3770)}$, we also extract
  $\Gamma_{ee}(\psi(3770))$.
  The experimentally observed cross section at 
  E$_{\mbox{\scriptsize{cm}}}=3773$~MeV 
  is related to the Born-level cross section by radiative 
  corrections~\cite{radker,radcor}, which, by convention, do not
  include vacuum polarization effects, allowing them to be absorbed
  into the definition of $\Gamma_{ee}$. These radiative effects account for
  virtual photon effects as well as real radiation down to lower energies 
  on the $\psi(3770)$ line-shape, effectively reducing the observed cross 
  section
  and amount to a reduction factor of $f=0.77\pm0.03$ based on
  the known mass $M$ and width $\Gamma$ of $\psi(3770)$ \cite{pdg}.
  The quoted uncertainty is dominated by the uncertainty in $M$ but
  also includes contributions from the uncertainties in 
  E$_{\mbox{\scriptsize{cm}}}$ (1.0~MeV)
  and in the phase space suppression of $D\bar{D}$ final states. 

  The Born-level cross section at the $\psi(3770)$ mass $M$ is related 
  to that at E$_{\mbox{\scriptsize{cm}}}=3773$~MeV via 
  the relativistic Breit-Wigner formula, 
  \begin{gather}
    \sigma_{\mbox{\scriptsize{Born}}}(\mbox{E}_{\mbox{\scriptsize{cm}}})
    =\frac{12\pi\Gamma_{ee}\Gamma_{\mbox{\scriptsize{total}}}}
    {(\mbox{E}_{\mbox{\scriptsize{cm}}}^2-M^2)^2
      +M^2\Gamma^2_{\mbox{\scriptsize{total}}}},
  \end{gather}
  which
  can be reduced to
  $\sigma_{\mbox{\scriptsize{Born}}}(M)
  /\sigma_{\mbox{\scriptsize{Born}}}
  (\mbox{E}_{\mbox{\scriptsize{cm}}})
  =1.078^{+0.152+0.055}_{-0.006-0.038}$ for
  $\mbox{E}_{\mbox{\scriptsize{cm}}}=3773$~MeV,
  in which the first error accounts mostly for the uncertainty in the
  PDG values for $M$ and $\Gamma$ and the
  second for the 1.0~MeV uncertainty in E$_{\mbox{\scriptsize{cm}}}$.

  The Breit-Wigner formula can then be used to extract
  $\Gamma_{ee}(\psi(3770))$ as 
  \begin{gather}
    \Gamma_{ee} = \frac{\sigma^{\mbox{\scriptsize{obs}}}
      (\mbox{E}_{\mbox{\scriptsize{cm}}})}{f}\times1.078\times M^2
    \times \Gamma_{\mbox{\scriptsize{total}}}/(12\pi).
  \end{gather}
  We obtain with PDG resonance parameters
  $\Gamma_{ee}(\psi(3770))=(0.204\pm0.003^{+0.041}_{-0.027})$~keV, where
  the first error is statistical and the second error is systematic including
  uncertainties of the input PDG values.
  The result is lower than, but consistent with and comparable in
  precision to, the
  PDG value of $0.26\pm0.04$~\cite{pdg}
  (the
  systematic errors there are mostly dominated by the uncertainties in $\Gamma$
  and $M$ of $\psi(3770)$).
  
  In summary, we have measured the hadronic cross section of $\psi(3770)$ at
  $\mbox{E}_{\mbox{\scriptsize{cm}}}=3773$~MeV,
  taking into account the effects of interference between 
  the final states of resonance decays and non-resonant annihilation of 
  $e^+e^-$ 
  with an improved relative uncertainty.
  The observed cross section is significantly smaller than some of the 
  previous measurements \cite{mark2,leadgas}. 
  By combining the reported cross section with that for 
  $\psi(3770)\to D\bar{D}$ \cite{xsecddcleo},
  we obtain $\sigma_{\psi(3770)}-\sigma_{\psi(3770)\to D\bar{D}}$.
  Based on the observed cross section of $\psi(3770)$, 
  we also extract $\Gamma_{ee}(\psi(3770))$.

  We gratefully acknowledge the effort of the CESR staff 
  in providing us with excellent luminosity and running conditions.
  This work was supported by 
  the A.P.~Sloan Foundation,
  the National Science Foundation,
  and the U.S. Department of Energy.

\end{document}